\documentclass[epj]{svjour}
\usepackage{graphics}

\begin{document}

\title{Wang-Landau study of the 3D Ising model with bond disorder}

\author{P.E. Theodorakis\inst{1,2,3}\thanks{e-mail:panagiotis.theodorakis@univie.ac.at} \and N.G. Fytas\inst{4}
\thanks{e-mail:nfytas@phys.uoa.gr}%
}                     

\institute{Faculty of Physics, University of Vienna,
Boltzmanngasse 5, 1090 Vienna, Austria \and Institute for
Theoretical Physics and Center for Computational Materials
Science, Vienna University of Technology, Hauptstra{\ss}e 8-10,
1040 Vienna, Austria \and Vienna Computational Materials
Laboratory, Sensengasse 8/12, 1090 Vienna, Austria \and Department
of Materials Science, University of Patras, 26504 Patras, Greece}

\date{Received: date / Revised version: date}

\abstract{We implement a two-stage approach of the Wang-Landau
algorithm to investigate the critical properties of the 3D Ising
model with quenched bond randomness. In particular, we consider
the case where disorder couples to the nearest-neighbor
ferromagnetic interaction, in terms of a bimodal distribution of
strong versus weak bonds. Our simulations are carried out for
large ensembles of disorder realizations and lattices with linear
sizes $L$ in the range $L=8-64$. We apply well-established
finite-size scaling techniques and concepts from the scaling
theory of disordered systems to describe the nature of the phase
transition of the disordered model, departing gradually from the
fixed point of the pure system. Our analysis (based on the
determination of the critical exponents) shows that the 3D
random-bond Ising model belongs to the same universality class
with the site- and bond-dilution models, providing a single
universality class for the 3D Ising model with these three types
of quenched uncorrelated disorder.
\PACS{
      {PACS. 05.50+q}{Lattice theory and statistics (Ising, Potts. etc.)}   \and
      {64.60.De}{Statistical mechanics of model systems} \and
      {75.10.Nr}{Spin-glass and other random models}
     }
}
\authorrunning{P.E. Theodorakis and N.G. Fytas} \titlerunning{Wang-Landau study of the 3D Ising model with bond disorder}

\maketitle

\section{Introduction}
\label{sec:1}

Understanding the role of impurities on the nature of phase
transitions is of great importance, both from experimental and
theoretical point of view~\cite{young}. First-order phase
transitions are known to be significantly softened under the
presence of quenched
randomness~\cite{aizenman-89,aizenman-89b,hui-89,hui-89b,chen-92,cardy-97,chatelain-98},
whereas continuous transitions may have their exponents altered
under random fields or random bonds~\cite{harris-74,chayes-86}.
There are some very useful phenomenological arguments and some,
perturbative in nature, theoretical results, pertaining to the
occurrence and nature of phase transitions under the presence of
quenched randomness~\cite{hui-89,hui-89b,dotsenko-95,jacobsen-98}.
Historically, the most celebrated criterion is that suggested by
Harris~\cite{harris-74}. This criterion relates directly the
persistence, under random bonds, of the non random behavior to the
specific heat exponent $\alpha_{p}$ of the pure system. According
to this criterion, if $\alpha_{p}>0$, then disorder will be
relevant, i.e., under the effect of the disorder, the system will
reach a new critical behavior. Otherwise, if $\alpha_{p}<0$,
disorder is irrelevant and the critical behavior will not change.
Pure systems with a zero specific heat exponent ($\alpha_{p}=0$)
are marginal cases of the Harris criterion and their study, upon
the introduction of disorder, has been of particular
interest~\cite{MK-99}. The paradigmatic model of the marginal case
is, of course, the general random 2D Ising model, which has been
extensively debated~\cite{gordillo-09}.

Respectively, the 3D Ising model with quenched randomness - which
is a clear case in terms of the Harris criterion having a positive
specific heat exponent in its pure version - has also been
extensively studied using Monte Carlo (MC)
simulations~\cite{landau-80,chow-86,heuer-90a,heuer-90b,heuer-93,hennecke-93,ball-98,wiseman-98,calabrese-03,hasenbusch-07,berche-04}
and field theoretical renormalization group
approaches~\cite{folk-00,pakhnin-00,pelissetto-00}. Especially,
the diluted model can be treated in the low-dilution regime by
analytical perturbative renormalization group
methods~\cite{newman-82,jug-83,mayer-89}, where a new fixed point
independent of the dilution has been found, yet for the strong
dilution regime only MC results remain valid. Although the first
numerical studies of the model suggested a continuous variation of
the critical exponents along the critical line, it soon became
clear that the concentration-dependent critical exponents found in
MC simulations are the effective ones characterizing the approach
to the asymptotic regime~\cite{heuer-90a,heuer-90b,heuer-93}.
Note, here, that a crucial problem of the new critical exponents
obtained in these studies is that the ratios $\beta/\nu$ and
$\gamma/\nu$ occurring in finite-size scaling (FSS) analysis are
almost identical for the disordered and pure models. In fact, for
the pure 3D Ising model, accurate values are~\cite{guida-98}:
$\nu=0.6304(13)$, $\beta/\nu=0.517(3)$, $\gamma/\nu=1.966(3)$, and
$\alpha=0.1103(1)$. Respectively, for the site- and bond-diluted
models, the most accurate sets of asymptotic exponents ($\nu$,
$\beta/\nu$, and $\gamma/\nu$) have been given by the numerical
works of Ballesteros et al.~\cite{ball-98} and Berche et
al.~\cite{berche-04} are: ($0.6837(53)$, $0.519(3)$, $1.963(5)$)
and ($0.68(2)$, $0.515(5)$, $1.97(2)$).

The above estimates of critical exponents provided evidence that
the 3D Ising model with quenched uncorrelated disorder belongs to
a single universality class, distinct from that of the pure model,
as also indicated by the Harris criterion, independent of the
considered disorder distribution. Yet, in a very recent paper
Murtazaev and Babaev~\cite{babaev-09} using MC simulations and FSS
methods on the site-diluted model, the above view was contradicted
and these authors suggested that the model has two regimes of
critical behavior universality, depending on the nonmagnetic
impurity concentration. Motivated by the above contradictions and
the great theoretical interest of the existence of universality
classes in disordered models, it seems favorable to investigate
the 3D Ising model with bond disorder in order to compare all
these three kinds of disorder (site-, bond-dilution, and bond
disorder) and to verify whether these lead to the same set of new
critical exponents, as would be, in principle, expected by
universality arguments~\cite{berche-04}. To this end,the first
approach has been recorded in a recent brief report by the present
authors~\cite{fytas-10}, who studied the random-bond model for a
single value of the disorder strength, obtaining values for the
critical exponents in very good agreement with the most accurate
estimates for the bond- and site-diluted models. Here, we extend
the analysis of reference~\cite{fytas-10}, including more values
of the disorder strength and several aspects of the FSS behavior
of the model, thus presenting a complete picture of the
disorder-induced phase transition of the 3D random-bond Ising
model. The main outcome of our work is that, indeed, the 3D Ising
model with quenched, uncorrelated bond disorder belongs to the
same universality class as the site- and bond-diluted models,
defining in this way a complete universality class in disordered
spin models.

The rest of the paper is organized as follows: In
Section~\ref{sec:2} we define the model by implementing the
bond-disorder distribution and describe the basic elements of our
two-stage numerical approach. In Section~\ref{sec:3} we present a
detailed FSS analysis of the obtained numerical data, estimate
critical exponents with high accuracy, and discuss the
disorder-induced second-order phase transition of the model under
the general prism of universality. This contribution closes with
Section~\ref{sec:4}, where a brief summary of our conclusions and
an outlook for further studies are given.

\section{Model and simulation method}
\label{sec:2}

We start this Section with the definition of the model. In the
following we consider the 3D bond-disorder Ising model, whose
Hamiltonian with uncorrelated quenched random interactions reads
\begin{equation}
\label{eq:1} H=-\sum_{\langle ij \rangle}J_{ij}S_{i}S_{j},
\end{equation}
where the spin variables $S_{i}$ take on the values $-1,+1$,
$\langle ij \rangle$ indicates summation over all nearest-neighbor
pairs of sites, and the ferromagnetic interactions $J_{ij}>0$
follow a bimodal distribution of the form
\begin{equation}
\label{eq:2}
P(J_{ij})=\frac{1}{2}~[\delta(J_{ij}-J_{1})+\delta(J_{ij}-J_{2})].
\end{equation}
In equation~(\ref{eq:2}) we set $J_{1}+J_{2}=2$, $J_{1}>J_{2}$,
and $r=J_{2}/J_{1}$ reflects the strength of the bond randomness.
Additionally, in the following, we fix as usual
$2k_{B}/(J_{1}+J_{2})=1$, to set the temperature scale ($k_{B}=1$
also for simplicity). The values of the disorder strength $r$
considered throughout this paper are the following: $r=0.75/1.25$,
$0.5/1.5$, and $0.25/1.75$.

Resorting to large scale MC simulations is often necessary,
especially for the study of the critical behavior of disordered
systems. It is also well known that for such complex systems
traditional methods become inefficient and thus in the last few
years several sophisticated algorithms, some of them are based on
entropic iterative schemes, have been proven to be very
effective~\cite{newman-99}. In the last few years we have used an
entropic sampling implementation of the Wang-Landau (WL)
algorithm~\cite{wang-01a,wang-01b} to study some
simple~\cite{malakis-04a,malakis-04b}, but also some more complex
systems~\cite{malakis-06,fytas-08a,fytas-08b,malakis-09,fytas-10b}.
One basic ingredient of this implementation is a suitable
restriction of the energy subspace for the implementation of the
WL algorithm. This was originally termed as the critical minimum
energy subspace restriction~\cite{malakis-04a,malakis-04b} and it
can be carried out in many alternative ways, the simplest being
that of observing the finite-size behavior of the tails of the
energy probability density function of the
system~\cite{malakis-04b}.

Complications that may arise in complex systems, i.e., random
systems or systems showing a first-order phase transition, can be
easily accounted for by various simple modifications that take
into account possible oscillations in the energy probability
density function and expected sample-to-sample fluctuations of
individual realizations. Recently, details of various
sophisticated routes for the identification of the appropriate
energy subspace $(E_{1},E_{2})$ for the entropic sampling of each
realization have been presented in
references~\cite{fytas-08a,fytas-08b}. To estimate the appropriate
subspace from a chosen pseudocritical temperature one should be
careful to account for the shift behavior of other important
pseudocritical temperatures and extend the subspace appropriately
from both low- and high-energy sides (as also discussed in
reference~\cite{fytas-08a}) in order to achieve an accurate
estimation of all finite-size anomalies. Of course, taking the
union of the corresponding subspaces, ensures accuracy for the
temperature region of all studied pseudocritical temperatures.

The up to date version of our implementation uses a combination of
several stages of the WL process. First, we carry out a starting
(or preliminary) multi-range (multi-R) stage, in a very wide
energy subspace. This preliminary stage is performed up to a
certain level of the WL random walk. The WL refinement is
$G(E)\rightarrow f\cdot G(E)$, where $G(E)$ is the density of
states (DOS) and we follow the usual modification factor
adjustment $f_{j+1}=\sqrt{f_{j}}$ and
$f_{1}=e$~\cite{malakis-04a,malakis-04b}. The preliminary stage
may consist of the levels : $j=1,\ldots,j=18$ and to improve
accuracy the process may be repeated several times. However, in
repeating the preliminary process and in order to be efficient, we
use only the levels $j=13,\ldots,18$ after the first attempt,
using as starting DOS the one obtained in the first random walk at
the level $j=12$. From our experience, this practice is almost
equivalent to simulating the same number of independent WL random
walks. Also in our recent studies we have found out that is much
more efficient and accurate to loosen up the originally applied
very strict flatness criteria~\cite{malakis-04a}. Thus, a variable
flatness process starting at the first levels with a very loose
flatness criteria and assuming at the level $j=18$ the original
strict flatness criteria is nowadays used. After the above
described preliminary multi-R stage, in the wide energy subspace,
one can proceed in a safe identification of the appropriate energy
subspace using one or more alternatives outlined in
reference~\cite{malakis-04b}.

\begin{figure}
\resizebox{1 \columnwidth}{!}{\includegraphics{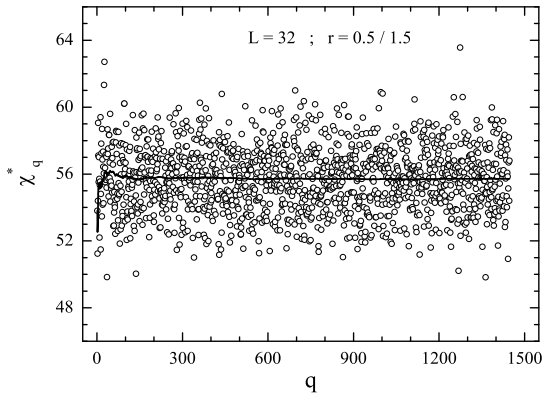}}
\caption{Disorder distribution of the susceptibility maxima of a
lattice with linear size $L=32$ for disorder strength $r=0.5/1.5$.
The running average over the samples is shown by the thick solid
line.} \label{fig:1}
\end{figure}

The process continues in two further stages (two-stage process),
using now mainly high iteration levels, where the modification
factor is very close to unity and there is not any significant
violation of the detailed balance condition during the WL process.
These two stages are suitable for the accumulation of histogram
data (for instance, energy-magnetization histograms), which can be
used for an accurate entropic calculation of non-thermal
thermodynamic parameters, such as the order parameter and its
susceptibility~\cite{malakis-04b}. In the first (high-level)
stage, we follow again a repeated several times (typically $\sim
5-10$) multi-R WL approach, carried out now only in the restricted
energy subspace. The WL levels may be now chosen as $j=18,19,20$
and as an appropriate starting DOS for the corresponding starting
level the average DOS of the preliminary stage at the starting
level may be used. Finally, the second (high-level) stage is
applied in the refinement WL levels $j=j_{i},\ldots,j_{i}+3$
(typically $j_{i}=21$), where we usually test both an one-range
(one-R) or a multi-R approach with large energy intervals. In the
case of the one-R approach we have found very convenient and more
accurate to follow the Belardinelli-Pereyra~\cite{belardinelli-07}
adjustment of the WL modification factor according to the rule
$\ln f\sim t^{-1}$. Finally, it should be also noted that by
applying in our scheme a separate accumulation of histogram data
in the starting multi-R stage (in the wide energy subspace) offers
the opportunity to inspect the behavior of all basic thermodynamic
functions in an also wide temperature range and not only in the
neighborhood of the finite-size anomalies.

\begin{figure}
\resizebox{1 \columnwidth}{!}{\includegraphics{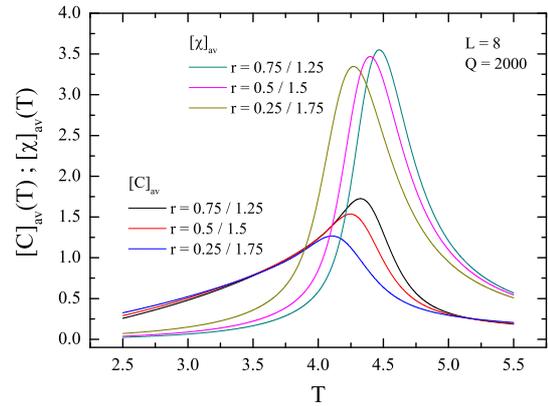}}
\caption{(Color online) Disorder-averaged specific heat $[C]_{av}$
(lower curves) and magnetic susceptibility $[\chi]_{av}$ (upper
curves) as a function of the temperature $T$ for a lattice with
linear size $L=8$. The ensemble of random realizations is $Q=2000$
for all cases of the disorder strength shown here. Notice the
expected shift of the pseudocritical temperatures to the left with
increasing the disorder strength ($r\rightarrow 0$).}
\label{fig:2}
\end{figure}

Using this scheme we performed extensive simulations for several
lattice sizes in the range $L=8-64$, over large ensembles
$\{1,\cdots,q,\cdots,Q\}$ of random realizations ($Q=1000 -
3000$). It is well known that, extensive disorder averaging is
necessary for the study of random systems, where usually broad
distributions are expected leading to a strong violation of
self-averaging~\cite{wiseman-98,aharony-96}. A measure from the
scaling theory of disordered systems, whose limiting behavior is
directly related to the issue of
self-averaging~\cite{wiseman-98,aharony-96} may be defined with
the help of the relative variance of the sample-to-sample
fluctuations of any relevant singular extensive thermodynamic
property $Z$ as follows:
$R_{Z}=([Z^{2}]_{av}-[Z]^{2}_{av})/[Z]^{2}_{av}$.
Figure~\ref{fig:1} presents evidence that the above number of
random realizations is sufficient in order to obtain the true
average behavior and not a typical one. In particular, we plot in
this figure (for a lattice size $L=32$ and disorder strength
$r=0.5/1.5$) the disorder distribution of the susceptibility
maxima $\chi_{q}^{\ast}$ (where the subscript $q$ denotes the
random realization) and the corresponding running average, i.e., a
series of averages of different subsets of the full data set -
each of which is the average of the corresponding subset of a
larger set of data points, over the samples for the simulated
ensemble of $Q=1447$ disorder realizations. A first striking
observation from this figure is the existence of very large
variance of the values of $\chi_{q}^{\ast}$, indicating the
violation of self-averaging for this
quantity~\cite{wiseman-98,aharony-96}. This figure illustrates
that the simulated number of random realizations is sufficient in
order to probe correctly the average behavior of the system, since
already for $Q\approx 300$ the average value of $\chi^{\ast}_{q}$
is stable.

Closely related to the above issue of self-averaging in disordered
systems is the manner of averaging over the
disorder~\cite{bernardet-00}. This non-trivial process may be
performed in two distinct ways when identifying the finite-size
anomalies, such as the peaks of the magnetic susceptibility. The
first way corresponds to the average over disorder realizations
($[\ldots]_{av}$) and then taking the maxima
($[\ldots]^{\ast}_{av}$), or taking the maxima in each individual
realization first, and then taking the average
($[\ldots^{\ast}]_{av}$). In the present paper we present our FSS
analysis using mainly the first approach of averaging, although we
should note that also the second gave comparable results. As an
example, we show in Figure~\ref{fig:2} the curves of the specific
heat and magnetic susceptibility for a lattice with linear size
$L=8$ averaged over $Q=2000$ random realizations. One may retrieve
the location of the pseudocritical point by taking the maximum of
these curves. Commenting on the statistical errors of our WL
scheme, we found that the statistical errors of our scheme on the
observed average behavior were of small magnitude (of the order of
the symbol sizes) and thus are neglected in the figures. On the
other hand, for the case $[\ldots^{\ast}]_{av}$ the error bars
shown reflect the sample-to-sample fluctuations.

\begin{figure}
\resizebox{1 \columnwidth}{!}{\includegraphics{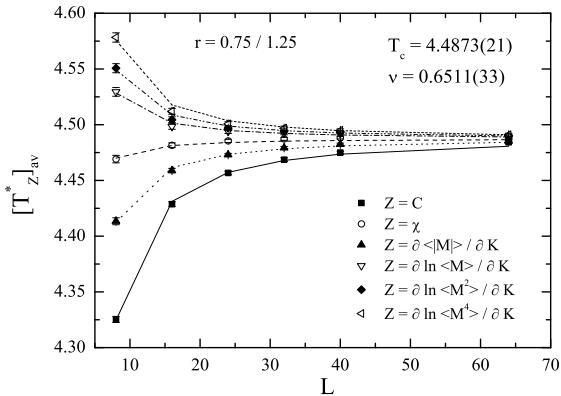}}
\caption{Shift behavior of several pseudocritical temperatures
defined in the text for $r=0.75/1.25$. The error bars reflect the
sample-to-sample fluctuations.} \label{fig:3}
\end{figure}

\section{Numerical results and discussion}
\label{sec:3}

\begin{figure}
\resizebox{1 \columnwidth}{!}{\includegraphics{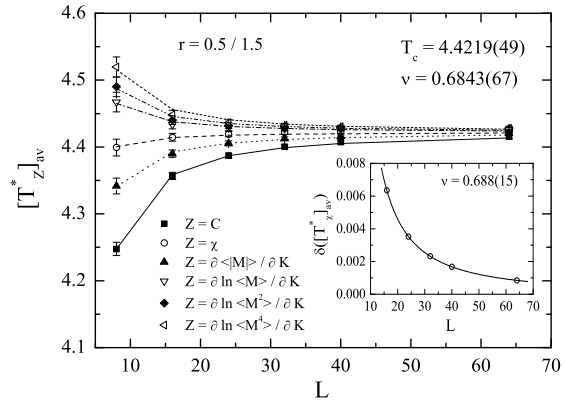}}
\caption{Same as Figure~\ref{fig:3}, but for $r=0.5/1.5$. The
inset shows the FSS of the sample-to-sample fluctuations of the
pseudocritical temperature of the magnetic susceptibility.}
\label{fig:4}
\end{figure}

\begin{figure}
\resizebox{1 \columnwidth}{!}{\includegraphics{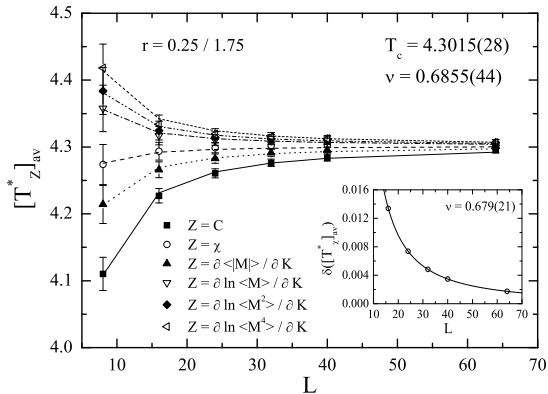}}
\caption{Same as Figure~\ref{fig:4} but for the case
$r=0.25/1.75$.} \label{fig:5}
\end{figure}

We present in this Section our numerical results and FSS analysis
for the 3D random-bond Ising model. In Figures~\ref{fig:3} -
\ref{fig:5} we illustrate in the main panels the shift behavior of
several pseudocritical temperatures, i.e., we take the average
over the pseudocritical temperatures of the individual samples. In
all cases the error bars reflect the sample-to-sample
fluctuations. The considered pseudocritical temperatures
correspond to the peaks of the following six quantities: specific
heat $C$, magnetic susceptibility $\chi$, derivative of the
absolute order parameter with respect to inverse temperature
$K=1/T$~\cite{ferrenberg-91}
\begin{equation}
\label{eq:3} \partial \langle |M|\rangle/\partial K=\langle
|M|H\rangle-\langle |M|\rangle\langle
H\rangle
\end{equation}
and logarithmic derivatives of the first ($n=1$), second ($n=2$),
and fourth ($n=4$) powers of the order parameter with respect to
inverse temperature~\cite{ferrenberg-91}
\begin{equation}
\label{eq:4}
\partial \ln \langle M^{n}\rangle/\partial K=\langle
M^{n}H\rangle/\langle M^{n}\rangle-\langle H\rangle.
\end{equation}
Fitting our data for the whole lattice range to the expected
power-law behavior
\begin{equation}
\label{eq:5} [T^{\ast}_{Z}]_{av}=T_{c}+bL^{-1/\nu},
\end{equation}
where $Z$ stands for the different thermodynamic quantities
mentioned above, we estimate the critical temperatures as a
function of the disorder strength and the respective critical
exponents of the correlation length. In particular, for the case
of very weak disorder (case $r=0.75/1.25$ shown in
Figure~\ref{fig:3}) we estimate a value $T_{c}=4.4873(21)$ for the
critical temperature and $\nu=0.6511(33)$ for the respective
critical exponent. Although this value of $\nu$ is larger than the
corresponding value of the pure system, it is still far away from
the expected value of the random system (as obtained in
references~\cite{ball-98,berche-04,fytas-10} and
Figures~\ref{fig:4} and \ref{fig:5}). This is due to finite-size
effects that dominate in the regime of very weak randomness,
hindering the approach of the system to the asymptotic limit. On
the other hand, the simultaneous fittings of the form~(\ref{eq:5})
shown in the main panels of Figures~\ref{fig:4} and \ref{fig:5}
for stronger values of the disorder strength $r$ reveal estimates
for the correlation length exponent $\nu$ around the value
$0.6849(50)$, in excellent agreement with the values $0.6837(53)$
and $0.68(2)$ given by Ballesteros et al.~\cite{ball-98} and
Berche et al.~\cite{berche-04} for the site- and bond-diluted
models. Additionally, the estimated values for the critical
temperatures, in all Figures~\ref{fig:3}, \ref{fig:4}, and
\ref{fig:5}, follow the usual reduction trend with increasing
disorder strength ($r \rightarrow 0$), as already shown in
Figure~\ref{fig:2} of Section~\ref{sec:2}.

Using the above sample-to-sample fluctuations of the
pseudocritical temperatures and the theory of FSS in disordered
systems as introduced by Aharony and Harris~\cite{aharony-96} and
Wiseman and Domany~\cite{wiseman-98}, one may further examine the
nature of the fixed point controlling the critical behavior of the
disordered system. According to the theoretical
predictions~\cite{aharony-96,wiseman-98}, the pseudocritical
temperatures $T_{Z}^{\ast}$ of the disordered system are
distributed with a width $\delta [T_{Z}^{\ast}]_{av}$, that
scales, in the case of a new random fixed point, with the system
size as
\begin{equation}
\label{eq:6} \delta([T_{Z}^{\ast}]_{av})\sim L^{-1/\nu}.
\end{equation}
In the insets of Figures~\ref{fig:4} and \ref{fig:5} we plot these
sample-to-sample fluctuations of the pseudocritical temperature of
the magnetic susceptibility. The solid lines show, in both cases,
a very good power-law fitting giving the values $0.688(15)$ and
$0.679(21)$ for the exponent $\nu$, which is also in very good
agreement with the values denoted above and obtained via the
classical shift behavior and the accurate estimates in the
literature~\cite{ball-98,berche-04}.

\begin{figure}
\resizebox{1 \columnwidth}{!}{\includegraphics{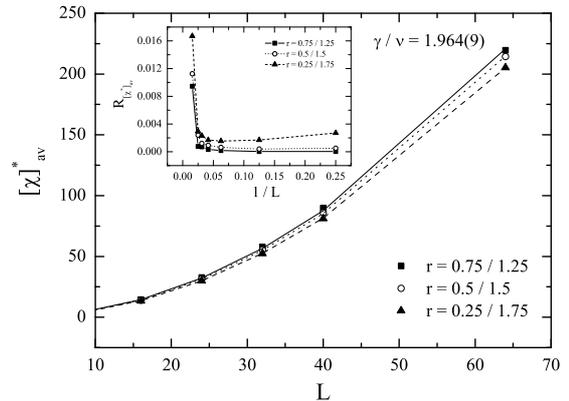}}
\caption{FSS of the maxima of the disorder-averaged magnetic
susceptibility for all the three values of the disorder strength
considered. The solid, dotted, and dashed lines show a
simultaneous fitting attempt of the form $[\chi]_{av}^{\ast}\sim
L^{\gamma/\nu}$ for the range $L=16-64$. The inset shows the
corresponding limiting behavior of the ratio
$R_{[\chi^{\ast}]_{av}}$ as defined in the text.} \label{fig:6}
\end{figure}

We move now to investigate the magnetic properties of the model.
In Figure~\ref{fig:6} we provide estimates for the magnetic
exponent ratio $\gamma/\nu$ of the 3D random-bond Ising model, by
presenting the FSS behavior of the maxima of the disorder-averaged
magnetic susceptibility $[\chi]_{av}^{\ast}$ for all the values of
the disorder strength considered. The solid, dotted, and dashed
lines show a simultaneous fitting of the form
\begin{equation}
\label{eq:7} [\chi]_{av}^{\ast}\sim L^{\gamma/\nu},
\end{equation}
using the $L=16-64$ lattice-range spectrum, providing an estimate
$1.964(9)$ for the exponent ratio $\gamma/\nu$. We remind the
reader that the respective fitting attempts of the numerical data
for each case separately gave the estimates $\gamma/\nu=1.962(4)$,
$1.967(3)$, and $1.969(4)$, for $r=0.75/1.25$, $0.5/1.5$, and
$0.25/1.75$, respectively. All these sets of values are in
excellent agreement with the accurate estimates $1.963(5)$ of
reference~\cite{ball-98} and $1.97(2)$ of
reference~\cite{berche-04} for the ratio $\gamma/\nu$ of the site-
and bond-diluted models, respectively. In the corresponding inset
of Figure~\ref{fig:6} we try to further quantify the behavior of
the sample-to-sample fluctuations of the model by plotting the
noise to signal ratio $R_{Z}$ - already introduced in
Section~\ref{sec:2} - as a function of the inverse lattice size,
for $Z=[\chi^{\ast}]_{av}$. Clearly, for the present model the
limiting value of $R_{[\chi^{\ast}]_{av}}$ is nonzero, indicating,
as expected also for marginal disordered
systems~\cite{wiseman-98}, a strong violation of self-averaging of
the magnetic properties of the 3D Ising model with bond disorder,
a property that intensifies with increasing disorder strength ($r
\rightarrow 0$). We should comment here that analogous behavior of
the ratio $R_{[\chi^{\ast}]_{av}}$ has been observed also in
several other 3D and 2D random models, some of them being the
bond-~\cite{berche-04} and site-diluted~\cite{ball-98} versions of
the present 3D Ising model and the random-bond versions of the
square Blume-Capel~\cite{malakis-09} and triangular
Ising~\cite{fytas-10b} model. As a final remark, we stress that
the asymptotic behavior of these $R$-ratios is expected, from
theoretical renormalization-group arguments, to follow specific
power law functions and reach limiting values independent of the
value of the disorder strength~\cite{aharony-96}. However, such an
analysis is by far much more demanding, in terms of numerical
data, as it has already been discussed by Berche et
al.~\cite{berche-04}, and, in any case, goes beyond the scope of
the present work.

\begin{figure}
\resizebox{1 \columnwidth}{!}{\includegraphics{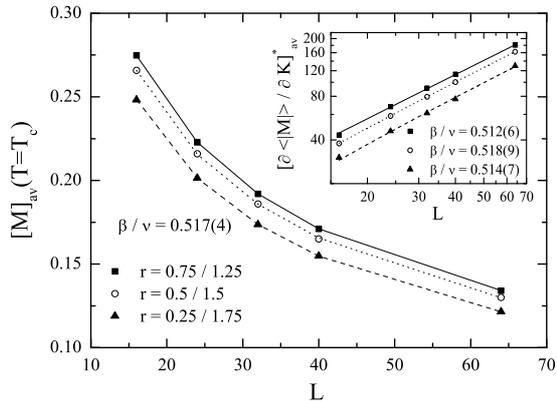}}
\caption{FSS of the disorder-averaged order parameter at the
estimated critical temperatures $T_{c}(r)$ for $r=0.75/1.25$,
$0.5/1.5$, and $0.25/1.75$. The inset illustrates the FSS of the
maxima of the disorder-averaged inverse-temperature derivative of
the absolute order parameter. In both main panel and inset,
lattices with linear sizes $L=16-64$ have been used in the
fittings.} \label{fig:7}
\end{figure}

Respectively, in Figure~\ref{fig:7} we plot the disorder-averaged
magnetization at the estimated critical temperatures of the
disordered model, as a function of the lattice size $L$. The
solid, dotted, and dashed lines show a simultaneous fitting
following the well-known relation
\begin{equation}
\label{eq:8} [M]_{av}(T=T_{c}) \sim L^{-\beta/\nu},
\end{equation}
giving an estimate of the order of $0.517(5)$ for the exponent
ratio $\beta/\nu$. Let us note here that, as in the case of the
magnetic susceptibility in Figure~\ref{fig:6}, the respective
separate fitting attempts of the numerical data gave comparable to
the $0.517$ estimates in the regime $0.511 - 0.522$. Additional
estimates for this exponent ratio can be obtained from the FSS of
the derivative of the absolute order parameter with respect to
inverse temperature $\partial \langle |M|\rangle/\partial K$.
Thus, in the corresponding inset of Figure~\ref{fig:7} we plot the
data for $\partial \langle |M|\rangle/\partial K$ averaged over
disorder as a function of $L$, in a double logarithmic scale.
Their maxima are expected to scale with the system size
as~\cite{ferrenberg-91}
\begin{equation}
\label{eq:9}  (\partial \langle |M|\rangle/\partial K)^{\ast}\sim
L^{(1-\beta)/\nu}.
\end{equation}
As in the main panel, the three lines shown in the inset
correspond to linear fittings for the three values of the disorder
strength, verifying the estimates of the main panel of the figure
for $\beta/\nu$, being also in very good agreement with accurate
estimates $0.519(3)$~\cite{ball-98} and
$0.515(5)$~\cite{berche-04} for the site- and bond-diluted models,
respectively.

Overall, the estimates for the critical exponent $\nu$ of the
correlation length and the magnetic exponent ratios $\gamma/\nu$
and $\beta/\nu$ given in this Section, and summarized in
Table~\ref{tab:1}, are in excellent agreement with accurate
estimates in the literature for the respective
site-~\cite{ball-98} and bond-diluted models~\cite{berche-04},
reinforcing the scenario of a single distinctive universality
class in the 3D Ising model with quenched uncorrelated disorder,
indicating the presence of a new random fixed point.

\section{Summary and outlook}
\label{sec:4}

In summary, we reported in the present paper the effects induced
by the presence of quenched bond randomness on the critical
behavior of the 3D Ising spin model embedded in the simple cubic
lattice, by implementing an efficient two-stage entropic
simulation scheme based on the Wang-Landau algorithm. Our
numerical approach enabled us to simulate large ensembles of
disorder realizations of the model for very large lattice sizes
(up to $N=64^{3}$) and several disorder strengths, and thus,
obtain, via finite-size scaling techniques, accurate estimates of
all the critical exponents describing the phase transition of the
model. Particular interest was paid to the sample-to-sample
fluctuations of the pseudocritical temperatures of the model and
their scaling behavior that was used as a successful alternative
approach to criticality.

The main conclusion of this study can be synopsized as follows:
the critical behavior of the 3D Ising model with quenched
uncorrelated disorder is controlled by a new random fixed point,
independently of the way randomness is implemented in the system.
It should be acknowledged at this point that, Ballesteros et
al.~\cite{ball-98} and Berche et al.~\cite{berche-04} were the
first scientific groups that strongly supported the above view for
the present model. However, we believe that the results brought
forward in this paper, combined with the existing ones for the
site- and bond-diluted models, complete the picture, at least for
this specific case of the 3D disordered Ising model, contributing
further to the understanding of the concept of universality in
disordered spin models.

\begin{table}
\caption{Summary of critical exponents for the pure and disordered
3D Ising model. The values of references~\cite{guida-98},
\cite{ball-98}, and \cite{berche-04} have been used for the pure,
site-, and bond-diluted versions of the model.} \label{tab:1}
\begin{tabular}{cccc}
\hline\hline\noalign{\smallskip}
Disorder Distribution &$\nu$&$\gamma/\nu$&$\beta/\nu$  \\
\noalign{\smallskip}\hline\noalign{\smallskip}
None (Pure model)&0.6304(13)&1.966(3)&0.517(3)\\
Site dilution&0.6837(53)&1.963(5)&0.519(3)\\
Bond dilution&0.68(2)&1.97(2)&0.515(5)\\
Bond randomness&0.685(7)&1.964(9)&0.517(4)\\
\noalign{\smallskip}\hline\hline
\end{tabular}
\end{table}

Closing, we stress here that it would be interesting to study
further the universality aspects of even more complicated spin
models, in terms of different disorder distributions and
couplings. One interesting case would be to consider
field-disorder coupling to the local order parameter, and one
prominent candidate for this case is the 3D random-field Ising
model~\cite{vink-10,sourlas-99}. Of course for this type of model
one may be particularly careful, since in this case there exist
the well-known hyperscaling violation and new scaling concepts
should be taken under serious
account~\cite{vink-10,sourlas-99,fytas-11}. Several attempts
towards this direction of understanding universality in random
systems are currently under consideration on both numerical and
theoretical grounds.

\end{document}